\documentstyle[aps,preprint,tighten,epsfig]{revtex}

\newcommand{\be}{\begin{equation}}
\newcommand{\ee}{\end{equation}}
\newcommand{\bea}{\begin{eqnarray}}
\newcommand{\eea}{\end{eqnarray}}
\newcommand{\no}{\noindent}
\newcommand{\bem}{\begin{mathletters}}
\newcommand{\eem}{\end{mathletters}}
\begin{document}

\draft

\title{Lepton Number Violation in Top-Color Assisted Technicolor}
\author{Tongu\c{c} Rador\thanks{E-mail: rador@buphy.bu.edu}} 
\address{Department of Physics, Boston University\\
590 Commonwealth Avenue, Boston, MA 02215}

\maketitle

\begin{abstract}
We calculate the rates for lepton number violating processes via
the exchange of  the $Z'$ boson occuring 
in top-color assisted technicolor. We found that 
$\mu\text{-}e$ conversion in nuclei is about an order of magnitude better 
than $\mu\to 3e$ for
constraining the magnitudes of the lepton mixing angles. 
The decay $\mu\to e\gamma$ 
yields much weaker bounds. The current experimental limits
allow for a mass of the new gauge boson around $1\;{\rm TeV}$ and
the magnitudes of the mixing angles turn out to lie 
roughly between the analogous elements of the 
CKM matrix and its square root. 
\end{abstract}


\section{Introduction}

Observation of lepton number violation would be 
one of the most spectacular evidences for deviations from 
the standard model.
Lepton mixing arises naturally in many of the extensions of the 
standard model. Here we consider the top-color assisted 
technicolor (TC2) scenario
introduced by Hill \cite{hill}. In TC2 
there exists an extra 
$U(1)$ group which breaks
at a higher energy than the electroweak breaking scale. The couplings 
of this $U(1)$ are generally not generation-universal, 
so there will be flavor-changing neutral current processes 
including the possibility for 
lepton number violation.

After the exposition of the theoretical framework in section II, 
we present the calculations of  
the rates for $\mu\to e\gamma$, $\mu\to 3e$, 
$\mu\text{-}e$ conversion in nuclei and the intrinsic dipole moments for 
leptons in a TC2 scenario
for which Chivukula and Terning \cite{chiv}
did a study on constraints from precision $Z$ data.

The decay rate for 
$\mu\to e\gamma$ does not provide so stringent a limit 
since it is a one loop process involving a photon vertex.
With the current data,  $\mu\text{-}e$ conversion in Ti gives limits that 
are roughly an order of magnitude 
better than $\mu\to 3e$. This will considerably improve 
with the proposed MECO experiment 
\cite{meco}. For the TC2 scenario we have considered, 
we found that the
current experimental limits allow for mixing angles of magnitude
ranging between the analogous elements of 
CKM matrix K and the elements of $\sqrt{K}$.

\section{Theory} 

In TC2, one has $SU(2)\times U(1)_{1}\times U(1)_{2}$  
as the electroweak gauge symmetry. The extra $U(1)$ will generate 
a new neutral gauge boson with a mass expected to be around 
$1\;{\rm TeV}$ \cite{1tev}. We require the breaking to occur in 
two stages, in the following pattern

{\Large
\[
{\underbrace{{}_{SU(2)\times}\underbrace{{}_{U(1)_{1}
\times U(1)_{2}}}_{U(1)_{Y}}}_{U(1)_{EM}}}
\]
}
\no Here $Y_{1}$ and $Y_{2}$ are the 
generators of $U(1)_{1}$ and $U(1)_{2}$ respectively 
and $Y=Y_{1}+Y_{2}$ is the ordinary hypercharge.

After the first step of symmetry breaking we want the gauge boson 
corresponding to $U(1)_{Y}$ to remain massless so that the second stage 
can proceed as in the standard model. This requires the first stage to be 
triggered by a neutral, $SU(2)$-singlet condensate. The second 
stage can be triggered by $SU(2)$-doublet condensates 
as in the standard model\footnote{
Here we adopt a Higgs-like formalism for the sake of future  
clarity in notation.
In TC2 fermions don't acquire masses via fundamental scalars. If
one insists on using fundamental scalars to generate fermion masses 
one will need more than two doublets to provide generational mixing 
via neutral currents of the new gauge boson since models with less than three
Higgs doublets are equipped with a natural GIM mechanism.}.

\no The covariant derivative for  $SU(2)\times U(1)_{1}\times U(1)_{2}$ 
is  $\partial^{\mu}+ig T^{a}W_{a}^{\mu}+ig'_{1}Y_{1}B_{1}^{\mu}+
ig'_{2}Y_{2}B_{2}^{\mu}$. Here $T^{a}\,\, , (a=1,2,3)$ are the generators 
of $SU(2)$. The gauge couplings can be parameterized as $g=e/\sin\theta$, 
$g'_{2}=e/\cos\theta\,\cos\phi$ and $g'_{1}=e/\cos\theta\,\sin\phi$. Here 
$\theta$ is the weak mixing angle and 
$\phi$ is a new mixing angle. The value of $\sin^{2}\phi$ should be smaller
than $1/2$ since this would mean interchanging the labeling of the $U(1)$'s. 
Furthermore, in TC2 one of the $U(1)$'s is strong, choosing it to be
$U(1)_{1}$, one has $\alpha_{2}={g'}_{1}^{2}/4\pi\simeq O(1)$ and this gives
roughly $\sin^{2}\phi\approx O(0.1)$. We now rotate the $B_{1}^{\mu}$, 
$B_{2}^{\mu}$ fields in terms of $\phi$ as follows

\bem
\bea{\label{eq:2}}
B^{\mu} &=& \cos\phi\, B_{2}^{\mu} + \sin\phi\, B_{1}^{\mu} \; , \\
Z_{2}^{\mu} &=& \cos\phi\, B_{1}^{\mu} - \sin\phi\, B_{2}^{\mu}\; . 
\eea
\eem

\no This choice of basis guarantees that $B^{\mu}$, coupling to $Y$, remains
massless after the first stage of the breaking pattern. The new gauge boson, 
$Z_{2}$, gets most of its mass
at this stage. 

For the second stage of symmetry breaking, we rotate the 
$B^{\mu}$ and $W_{3}^{\mu}$ fields in terms of $\theta$, as in the 
standard model,

\bem
\bea{\label{eq:3}}
A^{\mu} &=& \cos\theta\, B^{\mu} + \sin\theta W_{3}^{\mu} \; , \\
Z_{1}^{\mu} &=& \cos\theta\, W_{3}^{\mu} - \sin\theta\, B^{\mu}\; . 
\eea
\eem

\no The use of $SU(2)$-doublet condensates at this  stage 
provides mass to $Z_{1}$. The heavier boson, $Z_{2}$, also gets some 
additional mass and $A$ remains massless. 
The currents to which these gauge bosons couple, are as follows:

\bem
\bea
&A&\;\;\text{couples to}\;\;Q\equiv T^{3}+Y\;\;\text{with strength}\;\;e\; ,\\
&Z_{1}&\;\;\text{to}\;\;C\equiv T^{3}-Q\sin^{2}\theta\;\;
\text{with}\;\;g_{Z}\equiv 
e/\cos\theta\sin\theta\; , \\
&Z_{2}&\;\;\text{to}\;\;C'\equiv Y_{1}-Y\sin^{2}\phi\;\;\text{with}\;\; 
g_{Z'}\equiv e/\cos\theta\sin\phi\cos\phi\label{eq:dumbc} \; .
\eea
\eem

\noindent Generally, the mass matrix for these gauge bosons is not
diagonal, because the $SU(2)$-doublet condensates 
couple to both $Z_{1}$ and $Z_{2}$. Following the 
formalism in \cite{chiv} we write the mass eigenstates
as follows

\bem
{\label{eq:4}}
\bea
Z &\simeq& Z_{1} - \frac{\tan\phi\sin\theta}{\eta}(1+\frac{\xi}
{\sin^{2}\phi}) Z_{2} \; ,\\
Z' &\simeq&  \frac{\tan\phi\sin\theta}{\eta}(1+\frac{\xi}
{\sin^{2}\phi}) Z_{1} + Z_{2} \; . 
\eea
\eem

\no Here we introduced 

\bem
\bea{\label{eq:5}}
\xi &=& \sum_{_{j}}<T^{3}Y_{1}>_{_{j}}/\sum_{_{j}}<T^{3}T^{3}>_{_{j}} \;
,\\
\eta &=& \frac{\sin^{2}\theta}{\sin^{2}\phi\cos^{2}\phi}
\sum_{_{j}}<C'C'>_{_{j}}/\sum_{_{j}}<T^{3}T^{3}>_{_{j}} \; . 
\eea
\eem

\no We make use of Higgs language for the sake of notation; $j$ runs 
over the condensates used to trigger the breaking 
stages and $<X>_{_{j}}$ means the VEV of X with respect to the 
$j$'th condensate. Now, in natural TC2 models  the technifermion 
$Y_{2}$ hypercharges can be taken to be isospin symmetric \cite{isolane}. 
Then, since 
$Q=T_{3}+Y$ is conserved, the only contribution to $\xi$ comes from the
top-quark condensate. This is 

\be{\label{eq:xi}}
\xi=2\frac{f_{t}^{2}}{v^{2}}\left(Y_{1L}^{t}-Y_{1R}^{t}\right),
\ee

\no with $v\approx 250\;{\rm GeV}$ and $f_{t}\approx 64\;{\rm {GeV}}$ 
\cite{chiv} 
, the top-pion decay constant. Thus, 
$\xi\approx0.13(Y_{1L}^{t}-Y_{1R}^{t})$. 
The masses of the eigenstates in (\ref{eq:4}) are

\bem
\bea{\label{eq:6}}
M^{2}_{Z} & \simeq & M^{2}_{ZSM} \left(1-\frac{\tan^{2}\phi\sin^{2}\theta}
{\eta}(1+\frac{\xi}{\sin^{2}\phi})^{2}\right)\; ,\\
M^{2}_{Z'} & \simeq & \eta\;M^{2}_{ZSM}  \left(1+\frac{\tan^{2}\phi
\sin^{2}\theta}{\eta}(1+\frac{\xi}{\sin^{2}\phi})^{2}\right) \; .
\eea
\eem

\no Here $M_{ZSM}$ is the standard model prediction for the mass of $Z$. 
Then the correction to $\rho$ parameter due to the shift in the
$Z$ mass is given by

\be{\label{eq:rho}}
\delta\rho_{Z'} \simeq
\frac{\tan^{2}\phi\sin^{2}\theta}{\eta}(1+\frac{\xi}{\sin^{2}\phi})^{2}\; .
\ee

\no If $\xi=-\sin^{2}\phi$, there will be no $Z_{1}-Z_{2}$ mixing, but the
generation mixing effects of $Z_{2}$ will remain. The shift in $\rho$ 
must not be bigger than a percent \cite{PDG}. So, to this order, 
we must have

\be{\label{eq:eta}}
\eta\simeq\left(\frac{M_{Z'}}{M_{Z}}\right)^{2} \;.
\ee

\no In TC2, we expect $M_{Z'}\gtrsim 1-2\;{\rm TeV}$, meaning that
$\eta\gtrsim 100$.
\no We now rewrite the currents $Z$($Z'$) couple 
with $g_{Z}$($g_{Z'}$) factored out; 

\bem
\bea
J_{Z} &=& C-\zeta\sqrt{\delta\rho}\;C' \; ,\\
J_{Z'} &=& C'+\frac{g_{Z}^{2}}{g_{Z'}^{2}}\zeta\sqrt{\delta\rho}\;C \; .
\eea
\eem

\no Here, we suppressed chirality subscripts on $C$ and $C'$ for the sake 
of notation and omitted the $Z'$ subscript from $\delta\rho_{Z'}$. 
We also used $\zeta\equiv({g_{Z'}M_{Z}})/({g_{Z}M_{Z'}})\approx O(0.1)$ 
and $\delta\rho$
to have a compact and model independent notation.

If we consider $C$ and $C'$ as matrices in the 
3-dimensional generation space, $C$ is a 
multiple of the identity, since the standard model $Z$ has universal 
couplings. 
Thus, $C$ will commute with the rotation matrices 
that bring the leptons
to their mass eigenbasis.
However, there is no a priori reason for $C'$ to be a multiple of 
identity and in TC2 it isn't; after 
rotating the fermion fields, a non-universal $C'$
will induce tree-level generation mixing. We denote the rotated 
mass-eigenstate \footnote{Clearly, our motivation is independent 
of the fact that whether neutrinos have mass or not, because the tree-level
mixing is due to a neutral gauge boson.} charged lepton fields as 
follows (chiral indices are suppressed);

\be
\psi^{l} = \Lambda \psi'^{l} \; .
\ee

\no Here $\psi^{l}$, $(l=e,\mu,\tau)$, are the lepton mass eigenstates 
and $\Lambda$ is a $3\times3$ unitary matrix. 
As mentioned above $C'$ won't commute with the
rotation matrices, so we introduce the rotated lepton vertex matrices

\be{\label{eq:vertices}}
L\equiv\bar{\Lambda}C'^{l}\Lambda\; ,\label{eq:vertices1}
\ee

One of the biggest problems of such a theory is the flavor-changing neutral
currents involving the first two generations. This is cured in TC2 
by having the new gauge boson $Z_{2}$ couple with 
equal strength to the first two generations, and differently to the third. 
This implies the following

\bem
\bea
L^{e\mu}&=&(C'^{\tau}-C'^{e}){\bar{\Lambda}}^{e\tau}\Lambda^{\tau\mu}
\; ,\\
L^{l\tau}&=&(C'^{\tau}-C'^{e}){\bar{\Lambda}}^{l\tau}\Lambda^{\tau\tau}
\; , \;\;\;\; l=e,\mu \; .
\eea
\eem

\no Then, if one assumes 
$\Lambda^{\tau\tau}\gg\Lambda^{l\tau}$, $l=e,\mu$, 
one has

\be{\label{eq:iden1}}
L^{\mu\tau}\; , L^{e\tau} \gg L^{e\mu} \; .
\ee

\no Thus mixing between the first two generations is suppressed from the 
outset. 

In what follows, we will present the 
results for $\mu\to e\gamma$, the electron's electric dipole moment, 
$\mu\to 3e$ and $\mu\text{-}e$ conversion in Ti. To get a feel for the numerical
implications, we will use a TC2 model for which 
$Y_{1}=0$ ($C'=-Y\sin^{2}\phi$) 
for the first two generations and $Y_{1}=Y$ ($C'=Y(1-\sin^{2}\phi)$) 
for the third one. This results in $\xi\approx -0.07$. 
Chivukula and Terning \cite{chiv} fit the full precision $Z$ data 
including atomic parity violations to this model. 
The results of their fit are summarized in Fig.~\ref{fig1}. As can be 
seen from this graph, the lower bound for 
$M_{Z'}$ around $1-2\;{\rm Tev}$.

\section{Lepton Mixing Processes}
\subsection{{\bf The Amplitude for \boldmath{$ {\rm l} \to {\rm l'}\gamma$}}}

We calculated the amplitude for $l\to l'\gamma$ in a generalized $R_{\xi}$ 
gauge following the formalism of Ref.~\cite{bjlawe}. This process occurs at 
one loop ; the photon couples to the internal 
lepton propagator and the loop closes with a $Z$ or $Z'$ line. 
The matrix element is 
${\cal{M}}=\epsilon^{*}_{\mu}(q){\bar{u}}(p')\Gamma^{\mu}u(p)$ and 
$\Gamma^{\mu}$ is given by 

\be{\label{eq:ltolpg}}
\Gamma^{\mu}=-i\frac{eg_{Z'}^{2}}{16M_{Z'}^{2}\pi^{2}}\left[F_{_{+}}^{ll'}
\sigma^{\mu\nu}q_{\nu}-F_{_{-}}^{ll'}\sigma^{\mu\nu}q_{\nu}\gamma^{5}\right]\;.
\ee

\no With,

\be{\label{eq:ltolpg2}}
F_{_{\pm}}=\left[\frac{1}{3}\left[m,L^{2}_{L}\right]_{_{\pm}}-\frac{2}{3}
\frac{\sqrt{\delta\rho}}{\zeta}C_{L}\left[m,L_{L}\right]_{_{\pm}}-L_{L}mL_{R}
+\frac{\sqrt{\delta\rho}}{\zeta}C_{L}\left[m,L_{R}\right]_{_{\pm}}\right]
\;\pm\;\left({}_{L}\leftrightarrow {}_{R}\right)\;.
\ee

\no Here $\left[x,y\right]_{_{\pm}}=xy\pm yx$ and $m$ is the mass matrix of 
the leptons. From this amplitude one can calculate the decay $l\to l'\gamma$
and the electric dipole moments for $l$.

\subsubsection{{\bf The Electron Electric Dipole Moment}}

The electric dipole moments are given by the coefficient of 
$\sigma^{\mu\nu}q_{\nu}\gamma^{5}$ in Eq.~(\ref{eq:ltolpg}), 
so for the electron we 
need to evaluate $F^{ee}_{_{-}}$. This gives

\be{\label{eq:anomel2}}
d^{e}=\frac{e g_{Z'}^{2}}{16M_{Z'}^{2}\pi^{2}}{\rm Im}\left(m_{\tau}L_{L}^{e\tau}L_{R}^{\tau e} + m_{\mu}L_{L}^{e\mu}L_{R}^{\mu e}\right)\; .
\ee

Assuming Eq.~(\ref{eq:iden1}) holds, the RHS of the
equation above is dominated by the first term.  
The experimental value for $d^{e}$ is $(-0.27\pm 0.83)
\times 10^{-26} {\rm e\; cm}$ \cite{PDG}. Taking the 
$Z'$ contribution to lie within $1\sigma$, we get the 
following constraint:

\be{\label{eq:anomel3}}
M_{Z'}\gtrsim \frac{39.3}{\sin\phi\cos\phi}\left[{\rm Im}
\left({\bar{\Lambda}}_{L}^{e\tau}\Lambda_{L}^{\tau\tau}
{\bar{\Lambda}}_{R}^{\tau\tau}\Lambda_{R}^{\tau e}\right)\right]^{1/2}
\;\;{\rm TeV}.
\ee

\no If one considers the magnitude of the quantity above ignoring
the phases and assuming
$\Lambda_{L}\approx\Lambda_{R}\approx K$ one finds $M_{Z'}\gtrsim
0.14/(\sin\phi\cos\phi)\;\;{\rm TeV}$. Had we used
$\Lambda_{L}\approx\Lambda_{R}\approx \sqrt{K}$ this would change to
$M_{Z'}\gtrsim 0.02/(\sin\phi\cos\phi)\;\;{\rm TeV}$. Recalling that in
TC2 one expects $M_{Z'}$ to be around $1-2\;{\rm TeV}$, the former gives
$\sin^{2}\phi\gtrsim 0.01$ and the latter 
$\sin^{2}\phi\gtrsim 4\times 10^{-4}$.
These are expected, because very small values of $\sin^{2}\phi$ 
would make $g_{Z'}$ diverge and this in turn will result in a large mass for 
$Z'$.
On the other hand one can get rid of the electric dipole moments by assuming
$\Lambda_{R}^{i\tau}\approx\Lambda_{R}^{\tau i}\approx 0$, $i\neq\tau$, 
leaving $\Lambda_{L}$ 
unconstrained (or vice-versa). In the context of TC2, this type of behavior 
was strongly advocated for quark mixing angles to naturally eliminate the very 
stringent constraints resulting from $B^{o}_{d}-\bar{B^{o}}_{d}$ 
mixing \cite{komihill}.

\subsubsection{{\bf {\boldmath $\mu\to e\gamma$}}}

Using the amplitude in Eq.~(\ref{eq:ltolpg}) we find that the decay rate is

\be{\label{eq:mutoeg}}
\Gamma(\mu\to e\gamma)=\alpha_{e}\left(\frac{g_{Z'}^{4}m_{\mu}^{3}}
{2048\pi^{4}M_{Z'}^{4}}\right)\left[ |F_{_{+}}^{e\mu}|^{2} + 
|F_{_{-}}^{e\mu}|^{2}\right]\; .
\ee

With ${\rm BR}(\mu\to e\gamma)<4.9\times 10^{-11}$ \cite{PDG} and 
assuming for simplicity
$\Lambda_{R}^{i\tau}\approx\Lambda_{R}^{\tau i}\approx 0$,  $i\neq\tau$,
we have the following,

\be{\label{eq:mutoegc}}
\zeta^{2}|(C_{L}^{'\tau})^{2}-(C_{L}^{'e})^{2}|
\left|1-1.5\frac{\sqrt{\delta\rho}}{\zeta (C_{L}^{'\tau}+C_{L}^{'e})}\right| 
\lesssim \frac{7.2\times 10^{-4}}{\left|\Lambda_{L}^{\mu\tau}
\Lambda_{L}^{\tau e}\right|}\; .
\ee

\no Had we not used $\Lambda_{R}^{i\tau}\approx\Lambda_{R}^{\tau i}
\approx0,\;\; i\neq\tau$
the amplitude would
be dominated by $L_{L}mL_{R}$, which would make the RHS 
of Eq.~(\ref{eq:mutoegc})
smaller by an amount $m_{\mu}/m_{\tau}\simeq 0.06$. 
Even this will not help fix $Z'$ parameters; as we shall see shortly
other lepton number violating modes are better by orders of magnitude.

\subsection{\bf{ {\boldmath $\mu\to 3 e$}}}

This decay mode is allowed at tree level. One finds for the decay rate

\be{\label{eq:mto3e1}}
\Gamma(\mu\to3e)=\frac{m_{\mu}^{5}}{768\pi^{3}}\left(\frac{g_{Z'}}
{2M_{Z'}}\right)^{4}(3X+X')\; .
\ee

\no With $X$ and $X'$ given by,

\bem
\bea{\label{eq:mto3e}}
X&=&\left[|L_{V}^{e\mu}|^{2}+|L_{A}^{e\mu}|^{2}\right]
\left[(B_{V}^{ee})^{2}+(B_{A}^{ee})^{2}\right] \; ,\\
X'&=&2\left[L_{V}^{e\mu}(L_{A}^{e\mu})^{*}+L_{A}^{e\mu}
(L_{V}^{e\mu})^{*}\right]{B_{V}^{ee}}{B_{A}^{ee}} \; .
\eea
\eem

\no Here, we defined $B^{ee} \equiv  L^{ee}-\frac{\sqrt{\delta\rho}}{\zeta} 
C^{ee} = (B^{ee})^{*}$. Using ${\rm BR}(\mu\to 3e)<10^{-12}$ \cite{PDG} and 
assuming, for simplicity,  $\Lambda_{R}^{i\tau}\approx\Lambda_{R}^{\tau i}
\approx 0$, $i\neq\tau$, we have,

\be{\label{eq:mto3ec}}
\zeta^{2}|C_{L}^{'e}(C_{L}^{'\tau}-C_{L}^{'e})|\left|1-0.03
\frac{\sqrt{\delta\rho}}{\zeta C_{L}^{'e}}+0.03
\frac{\delta\rho}{\zeta^{2}(C_{L}^{'e})^{2}}\right|^{1/2} 
\lesssim \frac{2.2\times 10^{-7}}{{\left|\Lambda_{L}^{\mu\tau}
\Lambda_{L}^{\tau e}\right|}}\; .
\ee

\no 
This is by far a better constraint than the one imposed by $\mu\to
e\gamma$. The polynomial under the square root in (\ref{eq:mto3ec}) 
is stable between $1.00-1.06$ for $\sin^{2}\phi$ between $0.04-1$.
Thus, the constraint is dominated by the term
multiplying the square root. We compare this process with the $\mu\text{-}e$
conversion in Ti in the following subsection.

\subsection{\bf{ {\boldmath $\mu\text{-}e$} Conversion in Ti }}

This process is also a tree level process. Borrowing the result from
Barneb\'{e}u et.al. \cite{barnebeu}, one has the following rate
(normalized to the muon capture rate $\Gamma_{c}$):
\bem
\bea{\label{eq:mtoe}}
{\cal R}(\mu\text{-}e) &=& \frac{\alpha_{e}^{3} m_{\mu}^{5} 
Z_{eff}^{4} f^{2} g_{Z'}^{4}}{32\pi^{2} Z \Gamma_{c} 
M_{Z'}^{4}}\left[|L_{L}^{e\mu}|^{2}+|L_{R}^{e\mu}|^{2}\right] X^{2}\;, \\
X &=& (2Z+N) B_{V}^{uu}+(Z+2N) B_{V}^{dd} \;,\\
B^{qq} &\equiv& C'^{qq}-\frac{\sqrt{\delta\rho}}{\zeta} C^{qq} \;\; ,\; q=u,d.
\eea
\eem

\no The parameters for Ti are, $Z=22$, $N=26$, $Z_{eff}\simeq 17.6$,
$\Gamma_{c}\simeq 1.7\times 10^{-15}{\rm MeV}$ and $f\simeq 0.54$ 
\cite{barnebeu}. The current 
experimental upper bound ${\cal R}(\mu-e)_{Ti}<4.3\times 10^{-12}$ \cite{PDG} 
will give (again assuming for simplicity 
$\Lambda_{R}^{i\tau}\approx\Lambda_{R}^{\tau i}\approx0$, $i\neq\tau$) 

\be{\label{eq:mtoec}}
\zeta^{2}|C_{L}^{'e}(C_{L}^{'\tau}-C_{L}^{'e})|\left|1+0.14
\frac{\sqrt{\delta\rho}}{\zeta C_{L}^{'e}}\right|\lesssim 
\frac{3.6\times 10^{-8}}{{{\left|\Lambda_{L}^{\mu\tau}
\Lambda_{L}^{\tau e}\right|}}}
\ee

The term $\left|1+0.27{\sqrt{\delta\rho}}/{(\zeta C_{L}^{'e})}\right|$ 
is stable between $0.8-1$ for $\sin^{2}\phi$ between $0.04-1$, 
thus the constraint is dominated
by $\zeta^{2}|C^{'e}(C^{'\tau}-C^{'e})|$ as in the case of $\mu\to 3e$. 
So we see that RHS of 
(\ref{eq:mtoec}) is roughly 6 times the RHS of (\ref{eq:mto3ec}). 
This observation will remain roughly valid with the
relaxation of the assumption 
$\Lambda_{R}^{i\tau}=\Lambda_{R}^{\tau i}=0$, $i\neq\tau$. 
So we can  disregard the
process $\mu\to 3e$ and concentrate only on $\mu\text{-}e$ conversion for the
numerical analysis. Since the rate for $\mu\text{-}e$ conversion
depends on $|{L_{R}}^{e\mu}|^{2}+|{L_{R}}^{e\mu}|^{2}$ the correct form
of the constraint equation is

\be{\label{eq:mtoecc}}
\zeta^{2}|C_{L}^{'e}(C_{L}^{'\tau}-C_{L}^{'e})|
\left|1+0.14\frac{\sqrt{\delta\rho}}{\zeta C_{L}^{'e}}\right|\lesssim 
\frac{3.6\times 10^{-8}}{\delta} \;\; .
\ee

\no Here 

\be{\label{eq:mtoecc2}}
\delta \equiv {\sqrt{\left|\Lambda_{L}^{\mu\tau}\Lambda_{L}^{\tau
        e}\right|^{2}+4 \left|\Lambda_{R}^{\mu\tau}\Lambda_{R}^{\tau
        e}\right|^{2}}} \;\; .
\ee

\no The factor of $4$ in Eq.~(\ref{eq:mtoecc2}) is the ratio
$(Y_{R}/Y_{L})^{2}$
for leptons.

\section{Numerical analysis and conclusion}

In terms of the relevant quantities the constraint equation (\ref{eq:mtoecc})
reads

\be\label{eq:constf}
(\frac{{\rm TeV}}{M_{Z'}})^{2}\frac{1}{s^{2}c^{2}}|C_{L}^{'e}
(C_{L}^{'\tau}-C_{L}^{'e})|\left|1+3.2\frac{\sqrt{\delta\rho}}
{C_{L}^{'e}}(\frac{M_{Z'}}{{\rm TeV}})sc\right|\lesssim 
\frac{1.9\times 10^{-5}}{\delta}\;,
\ee

\begin{figure}
\centerline{\epsfig{figure=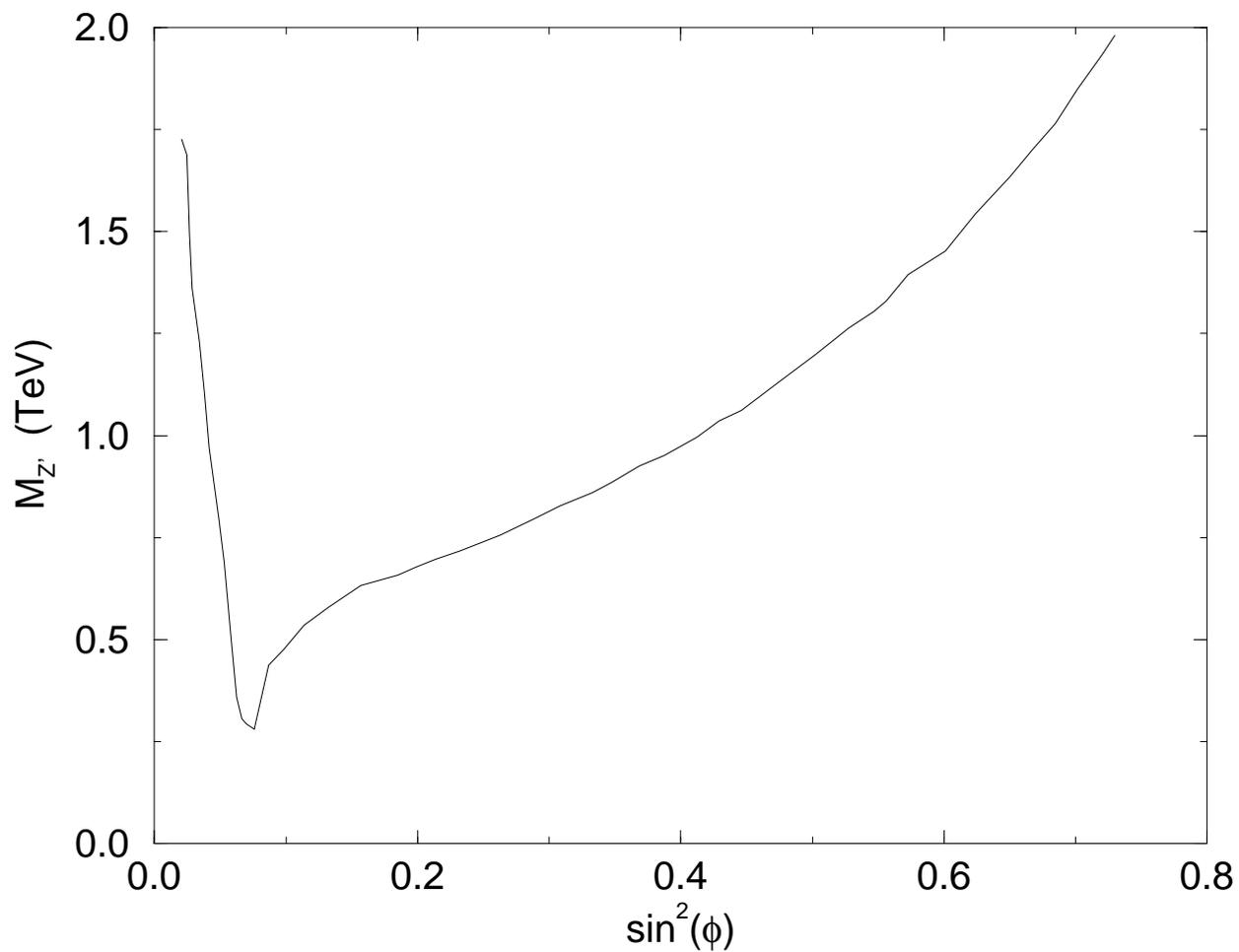}}
\caption{The $95\% $ confidence level lower bound on $M_{Z'}$ resulting from
the fit of the TC2 model we are considering to the precision $Z$ data. We 
reproduced the graph from Chivukula and Terning \protect\cite{chiv}.}
{\label{fig1}}
\end{figure}

\begin{figure}
\centerline{\epsfig{figure=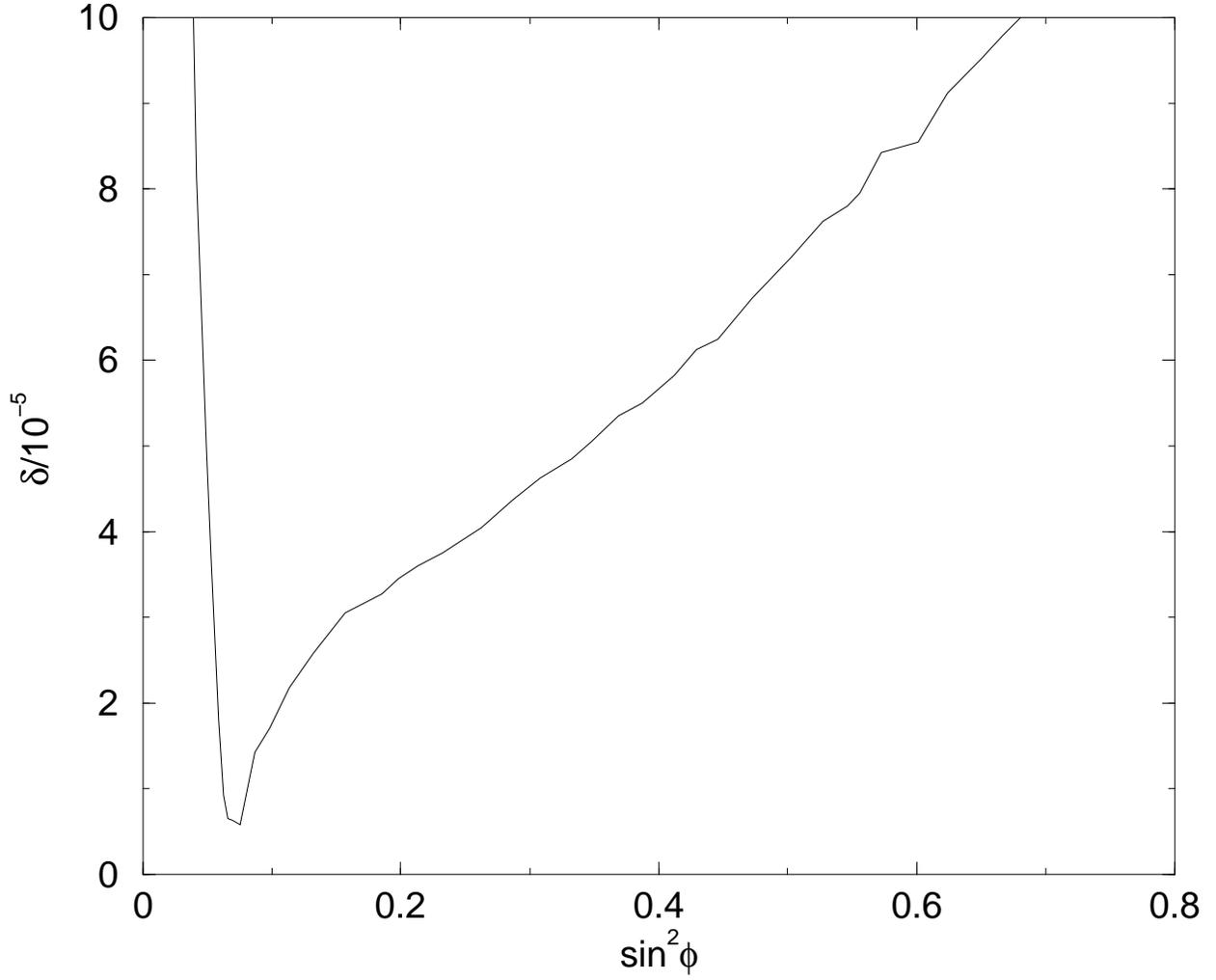}}
\caption{The upper bound on $\delta$, defined in Eq.~(\ref{eq:mtoecc2}), 
resulting from the  data in Fig.~\ref{fig1} and the constraint from 
$\mu- e$ conversion, Eq.~(\ref{eq:constf}).}
{\label{fig2}}
\end{figure}

\no where $s\equiv\sin\phi$ and $c\equiv\cos\phi$. Taking the lower 
limits for $M_{Z'}$ from Fig.~\ref{fig1} and feeding them
in the constraint equation for $\mu\text{-}e$ conversion, (\ref{eq:constf}), 
we get the upper limits on $\delta$ presented in Fig.~\ref{fig2}. 
As can be seen from Fig.~\ref{fig2} the lowest upper bound for $\delta$
occurs at $\delta\rho=0$, and it increases steeply for smaller values of
$\sin^{2}\phi$. This is because $g_{Z'}$ diverges for vanishingly 
small values of $\sin^{2}\phi$. The numerical upper bound for $\delta$
lies between $10^{-4}$ and $10^{-6}$ for the particular scenario
we have considered. These values are compatible with the naive estimates
made by taking $\Lambda\approx K$ or $\Lambda\approx \sqrt{K}$ for the
mixing matrices. The constraint eqution (\ref{eq:constf}) is also 
sensitive to the $Z'$ hypercharges: for example in the TC2 model
proposed recently by Lane \cite{lanelast} the bound is more stringent 
by a factor of $10$.

Since, within TC2, the reasons for expecting the $Z'$ mass 
around $1-2$ TeV are somewhat
robust, more stringent constraints may rule out lepton mixing
altogether. For example if the MECO experiment reaches the proposed
precision of $10^{-16}$ for $\mu\text{-}e$ conversion in Ti without observing
any candidate event, the upper bound on $\delta$ will be smaller by a
factor of $5\times 10^{-3}$. This will be hard to accomodate
with reasonable $Z'$ mass and hypercharges and will lead to the exclusion
of lepton number violation via $Z'$ for the TC2 model we have considered.

The main conclusion to be drawn is that the possibility of lepton
number violation in TC2 remains an interesting feature for now.
The proposed MECO experiment could tell which ones of the present models
involving lepton number violation may survive.

I thank Kenneth Lane for useful discussions and comments on the manuscript. 
This work was supported 
in part by the National Science Foundation under grant PHY-9501249,
and by the Department of Energy under grant DE-FG02-91ER40676.

\end{document}